\documentclass[preprint,prd,aps,nofootinbib,onecolumn,11pt]{revtex4}
\usepackage{amsmath}
\usepackage{epsfig}
\usepackage{amssymb}
\usepackage{color}

\begin{document}
	
\preprint{SI-HEP-2017-11}
\preprint{QFET-2017-08}

\def\dd{{$D^0$-$\overline{D}^0~$}}

\title{\dd mixing parameter $y$ in the factorization-assisted
topological-amplitude approach}
\author{Hua-Yu Jiang$^{1}$}
\author{Fu-Sheng Yu$^{1,2}$}\email{yufsh@lzu.edu.cn}
\author{Qin Qin$^{3,4}$}\email{qin@physik.uni-siegen.de}
\author{Hsiang-nan Li$^5$}\email{hnli@phys.sinica.edu.tw}
\author{Cai-Dian L\"u$^4$}\email{lucd@ihep.ac.cn}
\affiliation{$^1$School of Nuclear Science and Technology, Lanzhou University, Lanzhou 730000}
\affiliation{$^2$Research Center for Hadron and CSR Physics, Lanzhou University and Institute of Modern Physics of CAS, Lanzhou 730000}
\affiliation{$^3$Theoretische Physik 1, Naturwissenschaftlich-Technische Fakult\"at, Universit\"at Siegen, Walter-Flex-Strasse 3, D-57068 Siegen}
\affiliation{$^4$Institute of High Energy Physics, P.O. Box 918(4), Chinese Academy of Sciences, Beijing 100049}
\affiliation{$^5$Institute of Physics, Academia Sinica, Taipei 11529}

\begin{abstract}
We calculate the \dd mixing parameter $y$ in the
factorization-assisted topological-amplitude (FAT) approach, considering
contributions from $D^{0}\to PP$, $PV$, and $VV$ modes, where $P$ ($V$)
stands for a pseudoscalar (vector) meson. The $D^{0}\to PP$ and $PV$ decay
amplitudes are extracted in the FAT approach, and the $D^{0}\to VV$
decay amplitudes with final states in the longitudinal polarization are estimated via the
parameter set for $D^{0}\to PV$. It is found that the $VV$ contribution
to $y$, being of order of $10^{-4}$, is negligible, and that the $PP$ and
$PV$ contributions amount only up to $y_{PP+PV}=(0.21\pm0.07)\%$,
a prediction more precise than those previously obtained in the literature,
and much lower than
the experimental data $y_{\rm exp}=(0.61\pm0.08)\%$. We conclude that $D^{0}$
meson decays into other two-body and multi-particle final states are
relevant to the evaluation of $y$, so it is difficult to understand it fully in an exclusive approach.

\end{abstract}

\maketitle

\section{Introduction}
Studies of neutral meson mixings have marked glorious progress
in particle physics:  kaon mixing led to the first CP
violation observed in the $K_L\to\pi\pi$ decays \cite{GellMann:1955jx};
the masses of the charm quark \cite{Gaillard:1974hs} and top quark
\cite{Prentice:1987ap,Albajar:1986it} were, before their discoveries,
estimated through the GIM mechanism involved in kaon and $B_d$ meson
mixings, respectively. The neutral meson mixings are still a potential
regime for searching for new physics nowadays, because the relevant
flavor-changing  amplitudes are loop-suppressed in the Standard Model.
To get closer to this goal, it is crucial to understand the mixing
dynamics  to high precision.
The $B_{d(s)}$ meson mixing is well described in
the heavy quark effective theory \cite{Lenz:2006hd,Lenz:2011ti}, indicating
that both the power expansion parameter $1/m_b$ and the strong coupling
$\alpha_s(m_b)$ at the scale of the bottom quark mass $m_b$ are small
enough to justify a perturbative analysis. However,  understanding
\dd mixing has remained a challenge since its first observation
\cite{Aubert:2007wf,Staric:2007dt,Aaltonen:2007ac}.
It is suspected that $1/m_c$ and $\alpha_s(m_c)$, with $m_c$ being the charm
quark mass, may be too large to allow perturbative expansion.

The products $V_{ib}V^*_{id}$ of the Cabibbo-Kobayashi-Maskawa (CKM)
matrix elements, $i=u$, $c$, and $t$, which appear in the box diagram
responsible for the $B_{d}$ meson mixing, are of the same order. In the
$B_{s}$ meson mixing, $V_{tb}V^*_{ts}$ and $V_{cb}V^*_{cs}$
are of the same order, and both much larger than $V_{ub}V^*_{us}$.
Hence, an intermediate top quark with a much higher mass moderates the GIM
cancellation, giving a dominant contribution to the bottom mixing.
In the \dd mixing an intermediate bottom quark does not play
an important role due to the tiny product $V_{cb}V^*_{ub}$. The charm
mixing is then governed by the
difference between the other two intermediate quarks $s$ and
$d$, namely, by $SU(3)$ symmetry breaking effects, to which the
nonperturbative contribution is expected to be
significant.

The current world averages
of the charm mixing parameters are given by \cite{HFAG}
\begin{align}
x=(0.46^{+0.14}_{-0.15})\%,~~~~~y=(0.62\pm0.08)\%,
\end{align}
assuming no CP violation in charm decays
\footnote{As CP violation is allowed, the mixing parameters
turn into \cite{HFAG,Bevan:2014tha}
\begin{align}
x=(0.32\pm0.14)\%,~~~~~y=(0.69^{+0.06}_{-0.07})\%.
\end{align}}.
There are two approaches in the literature, inclusive and
exclusive, for the evaluation of the charm mixing
parameters. The former, with short-distance contributions calculated
based on the heavy quark expansion, leads to values of $x$ and $y$ two or three
orders of magnitude lower than the data, even after the
operators of dimension nine \cite{Bobrowski:2010xg,Bobrowski:2012jf} or both
$\alpha_s$ and subleading $1/m_c$ corrections \cite{Bobrowski:2010xg} are taken into account.
Obviously, the mass difference between the $s$ and $d$ quarks cannot collect all
$SU(3)$ breaking effects in charm decays, which may instead originate mainly
from hadronic final states \cite{Bigi:2000wn}. This speculation is supported by the argument
\cite{Jubb:2016mvq} that a modest quark-hadron duality violation of about 20$\%$
explains the discrepancy between inclusive predictions and the data.

Contributions to the charm mixing from individual intermediate
hadronic channels are summed up in an exclusive approach.
It was noticed \cite{Falk:2001hx,Falk:2004wg} that the $SU(3)$ breaking
effects only from the phase space naturally induce $x$ and $y$ at the order of
one percent, but are hard to predict quantitatively.
In a qualitative analysis based on $U$-spin and its breaking \cite{Gronau:2012kq},
it was found that contributions from two-body decays might be small,
and four-body decays may lead to $y$ at the measured level. The only
quantitative study in the literature was given in the topological
diagrammatic approach \cite{Cheng:2010rv}, showing that the $D \to P P$ and
$P V$ decays contribute to $y$ at the order of $10^{-3}$:
$y_{PP}=(0.86 \pm 0.41)\times10^{-3}$, $y_{PV}=(2.69\pm2.53)\times10^{-3} ~(A,A1)$
and $y_{PV}=(1.52 \pm2.20)\times10^{-3} ~(S,S1)$ from two different solutions.
The uncertainties of the predictions are too large to give a definite
conclusion. With abundant data collected on two-body $D$ meson decays
\cite{PDG}, it is now likely that a better control on $SU(3)$ breaking
effects can be obtained \cite{Li:2012cfa,Li:2013xsa}, and that the mixing parameter $y$ can be analyzed precisely  in an exclusive way.

In this paper we will address this issue in the factorization-assisted
topological-amplitude (FAT) approach \cite{Li:2012cfa,Li:2013xsa},
which provides a more precise treatment of the $SU(3)$ breaking effects
from two-body hadronic $D$ meson decays, as indicated by the improved global
fit to the measured branching ratios compared to Ref.~\cite{Cheng:2010rv}.
Distinct from the traditional diagrammatic approach based on
the $SU(3)$ symmetry \cite{Cheng:2010rv,Cheng:2010ry,Cheng:2010vk},
the $SU(3)$ breaking effects in phase space, decay constants, form factors,
and strong phases associated with various final states are captured in
the FAT approach. It is well known that the $SU(3)$ breaking effects in the
singly Cabibbo-suppressed modes are significant. For instance, the
ratio of the $D^0\to K^+K^-$ and $\pi^+\pi^-$ branching fractions should
be unity in the limit of $SU(3)$ symmetry, but is measured to be about 2.8.
This approach has been successfully applied to studies of the $D\to PP$
\cite{Li:2012cfa} and $D\to PV$ \cite{Li:2013xsa,Qin:2014nxa} decays,
including all the Cabibbo-favored, singly Cabibbo-suppressed,
and doubly Cabibbo-suppressed modes, as well as the
charmed \cite{Zhou:2015jba} and charmless \cite{Zhou:2016jkv,Wang:2017hxe} $B$
meson decays. In particular, the
predicted difference of the direct CP asymmetries $\Delta a_{\rm CP}^{\rm dir}\equiv
a_{\rm CP}^{\rm dir}(K^+K^-)-a_{\rm CP}^{\rm dir}(\pi^+\pi^-)=(-0.6\sim-1.9)\times 10^{-3}$
was later confirmed by the LHCb data, $\Delta a_{\rm CP}^{\rm dir}=(-0.61\pm0.76)\times10^{-3}$
\cite{Aaij:2016cfh}.

It is expected that the contributions from two-body $D$ meson decays
to the $D^0-\overline D^0$ mixing can be properly addressed in the FAT approach.
The $D^{0}\to PP$ and $PV$ decay amplitudes required for the evaluation of
the mixing parameter $y$ are extracted in the FAT approach. The
$D^{0}\to VV$ decay amplitudes with final states in the longitudinal polarization are estimated
via the parameter set for $D^{0}\to PV$, which does yield  corresponding
branching ratios in agreement with data. We will show that the
$D^0\to PP$, $PV$ and $VV$ channels contribute $y_{PP}=(0.10\pm0.02)\%$,
$y_{PV}=(0.11\pm0.07)\%$, and $y_{VV}\sim 10^{-4}$, respectively, to the mixing
parameter, with small uncertainties. Namely, the above two-body
channels alone, which take up about 50\% of the total $D^{0}$ meson
decay rate, cannot explain the \dd mixing in an exclusive approach.
Therefore, other two-body and multi-particle hadronic $D$ meson
decays are relevant to the calculation of $y$. These are, however, extremely
difficult to analyze in an exclusive approach at the current stage.
A new strategy to understand  charm mixing dynamics is necessary.

In Section~2 we update the determination of the $D^0\to PP$ and $PV$
amplitudes by performing a global fit to the latest data of the
branching ratios in the FAT approach. Their contributions to
the charm mixing parameter $y$ are then obtained. The $D^0\to VV$
amplitudes for the longitudinal polarization are estimated via the parameter
set for the $PV$ modes in Section~3, and found to give a small contribution to
$y$. Section 4 contains the summary.

\section{$y_{PP}$ and $y_{PV}$}

The \dd mixing parameter $y$ is defined by
\begin{align}
y\equiv {\Gamma_{1}-\Gamma_{2}\over 2\Gamma},
\end{align}
where $\Gamma_{1,2}$ represent the widths of the mass eigenstates $D_{1,2}$,
and $\Gamma=(\Gamma_{1}+\Gamma_{2})/2$. In the assumption
of CP conservation, the mass eigenstates are identical to the CP
eigenstates, {\it i.e.}, $|D_{1}\rangle=|D_{+}\rangle$ and $|D_{2}\rangle=|D_{-}\rangle$,
with $|D_{\pm}\rangle = ( |D^{0}\rangle \pm |\overline {D}^{0}\rangle )/\sqrt{2}$.
Here we adopt the convention of
$\mathcal{CP}|D^0\rangle = + |\overline{D}^0\rangle$. The parameter
$y$ can be computed via the formula
\begin{align}
y&={1\over2\Gamma}\sum_{n}\rho_{n}\left(|\mathcal{A}(D_{+}\to n)|^{2}
 - |\mathcal{A}(D_{-}\to n)|^{2}\right)
\nonumber\\
&={1\over\Gamma}\sum_{n}\eta_{\rm CP}(n)\rho_{n}\mathcal{R}e
\left[\mathcal{A}(D^{0}\to n)\mathcal{A}^{*}(D^{0}\to \bar n)\right],
\label{eq:yamp}
\end{align}
in which $\rho_{n}$ is the phase-space factor for the $D^{0}/\overline D^{0}$
decay into the final state $n$, and the transformation
$\mathcal{CP}|n\rangle=\eta_{\rm CP}|\bar n\rangle$ has been applied.
For the $PP$ and $PV$ modes, $\eta_{\rm CP}=+1$, and for the $VV$ modes,
$\eta_{\rm CP} = (-1)^L$, with $L$ denoting the
orbital angular momentum of the final state.
The following expression is also employed in the literature \cite{Cheng:2010rv},
\begin{align}\label{eq:ybr}
y=\sum_{n}\eta_{\rm CKM}(n) \eta_{\rm CP}(n) \cos\delta_{n}
\sqrt{\mathcal{B}(D^{0}\to n)\mathcal{B}(D^{0}\to \bar n)},
\end{align}
where $\delta_{n}$ is the relative strong phase between the $D^{0}\to n$ and
$D^{0}\to \bar n$ amplitudes, and $\eta_{\rm CKM}=(-1)^{n_{s}}$, with $n_{s}$
being the number of $s$ or $\bar s$ quarks in the final state.

The FAT approach is based on the factorization of short-distance and long-distance
dynamics in the topological amplitudes for $D$ meson decays into Wilson
coefficients and hadronic matrix elements of effective operators, respectively.
The relevant tree-level topological amplitudes include
the color-favored tree-emission diagram $T$, the color-suppressed
tree-emission diagram $C$, the $W$-exchange diagram $E$, and the
$W$-annihilation diagram $A$. The hadronic matrix elements are partly computed
in the naive factorization with nonfactorizable contributions being
parameterized into strong parameters. A $D$ meson decay amplitude is then
decomposed into these topological diagrams, each of which further takes into
account channel-dependent $SU(3)$ symmetry breaking effects. Through a global
fit to the abundant decay-rate data, the strong parameters are determined and can
be used to make predictions for unmeasured branching ratios and CP asymmetries.
The resultant channel-dependent phases will be employed for the evaluation of $y$ here.
It is noticed that the $W$-exchange diagram $E$ appears only in $D^{0}$ meson
decays, while the $W$-annihilation diagram $A$ contributes only to $D^{+}$ and
$D_{s}^{+}$ meson decays. For the study of  \dd mixing, we focus on the
$D^{0}$ meson decay modes, so that the irrelevant strong parameters associated
with the amplitudes $A$ can be removed from the global fits.

For the explicit parametrizations of the $D\to PP$ and $PV$ amplitudes in the FAT
approach, we refer to Refs.~\cite{Li:2012cfa} and \cite{Li:2013xsa}, respectively.
Below we update the sets of strong parameters
determined by the latest data:
\begin{align}\label{eq:paraPP}
&~~~~~~~\chi^{C}=-0.81\pm0.01,~~~~~\phi^{C}=0.22\pm0.14,~~~~~S_{\pi}=-0.92\pm0.07,
\nonumber\\
&\chi_{q}^{E}=0.056\pm0.002,~~~\phi_{q}^{E}=5.03\pm0.06,~~~\chi_{s}^{E}=0.130\pm0.008,~~~\phi_{s}^{E}=4.37\pm0.10,
\end{align}
for the $D^{0}\to PP$ decays, and
\begin{align}\label{eq:paraPV}
&S_{\pi}=-1.88\pm0.12,~~~~\chi_{P}^{C}=0.63\pm0.03,~~~~\phi_{P}^{C}=1.57\pm0.11,
\nonumber\\
&\chi_{V}^{C}=0.71\pm0.03,~~~~~~\phi_{V}^{C}=2.77\pm0.10,~~~~~~\chi_{q}^{E}=0.49\pm0.03,
\\
&\phi_{q}^{E}=1.61\pm0.07,~~~~~~\chi_{s}^{E}=0.54\pm0.03,~~~~~~\phi_{s}^{E}=2.23\pm0.08,~~~~\nonumber
\end{align}
for the $D^{0}\to PV$ decays. In both the $PP$ and $PV$ modes, the
parameter $\Lambda$ related to the soft scale in $D$ meson decays
is fixed to be 0.5 GeV. The decay constants of the vector mesons are from
Ref.~\cite{Straub:2015ica}, and other theoretical inputs are the same as in
Refs.~\cite{Li:2012cfa,Li:2013xsa}. The minimal $\chi^{2}$ per degree of
freedom is 1.1 for the data of 13 PP modes, and 1.8 for the data of 19 $PV$
modes. The $D^0 \to PP$ and $PV$ branching
fractions predicted in the FAT approach are given in Tables \ref{tablePP}
and \ref{tablePV}, and agree well with the data. The cosines of the
relative strong phases, $\cos\delta_n$, in Eq.~(\ref{eq:ybr}),
listed in the rows of the $D^0\to n$ and $D^0\to \bar n$
decays, reveal the channel dependence and the $SU(3)$ symmetry
breaking effects. Those shown as $1$ are for the modes with $CP$ eigenstates,
i.e., $n=\bar n$. Those shown as $1.0\pm0.0$ are for the modes, in which
the relative strong phases vanish with tiny uncertainties
in the FAT approach. The expression $1.0\pm0.0$ means that those 
strong phases can deviate from zero in principle, but turn out to 
vanish with tiny uncertainties in the FAT approach. The values of 
$\cos\delta_n$ can never be greater than unity. It is observed that the $D^0\to K^\pm\rho^\mp$
and $D^0\to K^\pm K^{*\mp}$ decays exhibit nonvanishing relative strong phases
around 10 degrees, different from the approximation $\cos\delta_{n}=1$
assumed in Ref.~\cite{Cheng:2010rv}. To confirm that the values of $\cos\delta_n$
are close to unity in the $D^0\to PP$ decays, we have allowed the $W$-exchange
diagrams $E$ in the Cabibbo-favored and doubly Cabibbo-suppressed modes to carry
different strong phases, which may lead to nonvanishing $\delta_n$.
The associated global fit indeed indicates that the results in Table~\ref{tablePP}
remain unaltered.

\begin{table}[!tb]
\caption{ Branching ratios in units of $10^{-3}$ and
 cosines of the relative strong phases for the $D^{0}\to PP$ decays.
Predictions $\mathcal{B}$(FAT) in the FAT approach are compared with the experimental
data $\mathcal{B}$(exp) \cite{PDG}.
Topological parametrizations are also given with $\lambda_{ij}=V_{ci}^*V_{uj}$, in which
each topological amplitude, including the $SU(3)$ symmetry breaking effects, is
actually mode-dependent.}\label{tablePP}
\begin{tabular}{cccccccccc}
\hline\hline
Modes & Parametrization & $\mathcal{B}$(exp) & $\mathcal{B}$(FAT) &  $\cos\delta_n$
\\\hline
$\pi^{0}\overline{K}^{0}$ & ${1\over\sqrt2}\lambda_{sd}(C-E)$ & $24.0\pm0.8$&$24.2\pm0.8$ & $1.0\pm0.0$
\\
$\pi^{+}K^{-}$ & $\lambda_{sd}(T+E)$ & $39.3\pm0.4$&$39.2\pm0.4$ &$0.99999\pm0.00001$
\\
$\eta\overline{K}^{0}$ & $\lambda_{sd}[{1\over\sqrt2}(C+E)\cos\phi - E\sin\phi ]$ & $9.70\pm0.6$&$9.6\pm0.6$ &$1.0\pm0.0$
\\
$\eta^{\prime}\overline{K}^{0}$ & $\lambda_{sd}[{1\over\sqrt2}(C+E)\sin\phi + E\cos\phi ]$ & $19.0\pm1.0$&$19.5\pm1.0$ &$1.0\pm0.0$
\\
\hline
 $\pi^{+}\pi^{-}$&$\lambda_{dd}(T+E)$&$1.421\pm0.025$&$1.44\pm0.02$&1
\\
 $ K^{+}K^{-}$&$\lambda_{dd}(T+E)$&$4.01\pm0.07$&$4.05\pm0.07$  &1
\\
 $ K^{0}\overline{K}^{0}$&$\lambda_{dd}E+\lambda_{ss}E$&$0.36\pm0.08$&$0.29\pm0.07$      &1
\\
 $\pi^{0}\eta$&$-\lambda_{dd}E \cos\phi-\frac{1}{\sqrt{2}}\lambda_{ss}C \sin\phi$&$0.69\pm0.07$&$0.74\pm0.03$ &1
\\
 $\pi^{0}\eta^{\prime}$&$-\lambda_{dd}E \sin\phi+\frac{1}{\sqrt{2}}\lambda_{ss}C \cos\phi$&$0.91\pm0.14$&1.08$\pm$0.05 &1
\\
 $\eta\eta$&$\frac{1}{\sqrt{2}}\lambda_{dd}(C+E)\cos^{2}\phi+\lambda_{ss}(2E \sin^{2}\phi-\frac{1}{\sqrt{2}}C \sin2\phi)$&$1.70\pm0.20$&1.86$\pm$0.06 &1
\\
 $\eta\eta^{\prime}$&$\frac{1}{\sqrt{2}}\lambda_{dd}(C+E)\sin2\phi+\lambda_{ss}(E \sin2\phi-\frac{1}{\sqrt{2}}C \cos2\phi)$&$1.07\pm0.26$&1.05$\pm$0.08 &1
\\
 $\pi^{0}\pi^{0}$&$\frac{1}{\sqrt{2}}\lambda_{dd}(C-E)$&$0.826\pm0.035$&$0.78\pm0.03$ &1
\\
\hline
 $\pi^{0}K^{0}$& ${1\over\sqrt2}\lambda_{ds}(C-E)$ & &0.069$\pm$0.002&$1.0\pm0.0$
\\
 $\pi^{-}K^{+}$&$\lambda_{ds}(T+E)$&$0.133\pm0.009$&0.133$\pm$0.001&$0.99999\pm0.00001$
\\
 $\eta K^{0}$&$\lambda_{ds}[\frac{1}{\sqrt{2}}(C+E)\cos\phi-E \sin\phi]$&&0.027$\pm$0.002&$1.0\pm0.0$
\\
 $\eta^{\prime} K^{0}$&$\lambda_{ds}[\frac{1}{\sqrt{2}}(C+E)\sin\phi+E \cos\phi]$&&0.056$\pm$0.003&$1.0\pm0.0$
\\
\hline \hline
\end{tabular}
\end{table}
\begin{table}[!tb]
\caption{Branching ratios in units of $10^{-3}$ and
	cosines of the relative strong phases for the $D^{0}\to PV$  decays.
	Predictions $\mathcal{B}$(FAT) in the FAT approach are compared with the experimental
	data $\mathcal{B}$(exp) \cite{PDG}.
	Topological parametrizations are also given with $\lambda_{ij}=V_{ci}^*V_{uj}$, in which
	each topological amplitude, including the $SU(3)$ symmetry breaking effects, is
	actually mode-dependent.}\label{tablePV}
\begin{tabular}{cccccccccc}
\hline\hline
Modes & Parametrization & $\mathcal{B}$(exp) & $\mathcal{B}$(FAT) &  $\cos\delta_n$
\\\hline
 $\pi^{0}\overline{K}^{\ast0}$ & ${1\over\sqrt2}\lambda_{sd}(C_P-E_P)$ & $37.5\pm2.9$&$35.9\pm2.2$ &$1.0\pm0.0$
\\
 $\overline{K}^{0}\rho^{0}$ & ${1\over\sqrt2}\lambda_{sd}(C_V-E_V)$ & $12.8^{+1.4}_{-1.6}$&$13.5\pm1.4$ &$1.0\pm0.0$
\\
 $\pi^{+}K^{\ast-}$ & $\lambda_{sd}(T_V+E_P)$ &$54.3\pm4.4$&$62.5\pm2.7$ &$0.9994\pm0.0006$
\\
 $K^{-}\rho^{+}$&$\lambda_{sd}(T_{P}+E_{V})$&$111.0\pm9.0$&$105.0\pm5.2$ &$0.983\pm0.002$
\\
 $\eta\overline{K}^{\ast0}$&$\lambda_{sd}(\frac{1}{\sqrt{2}}(C_{P}+E_{P})\cos\phi-E_{V}\sin\phi)$&$9.6\pm3.0$&$6.1\pm1.0$ &$1.0\pm0.0$
\\
 $\eta^{\prime}\overline{K}^{\ast0}$&$\lambda_{sd}(\frac{1}{\sqrt{2}}(C_{P}+E_{P})\sin\phi+E_{V}\cos\phi)$&$<1.10$&$0.19\pm0.01$ &$1.0\pm0.0$
\\
$\overline{K}^{0}\omega$&$\frac{1}{\sqrt{2}}\lambda_{sd}(C_{V}+E_{V})$&$22.2\pm1.2$&$22.3\pm1.1$ &$1.0\pm0.0$
\\
 $\overline{K}^{0}\phi$&$\lambda_{sd}E_{P}$&$8.47_{-0.34}^{+0.66}$&$8.2\pm0.6$ &$1.0\pm0.0$
\\
\hline
 $\pi^{+}\rho^{-}$&$\lambda_{dd}(T_{V}+E_{P})$&$5.09\pm0.34$&$4.5\pm0.2$ &$0.9995\pm0.0005$
\\
 $\pi^{-}\rho^{+}$&$\lambda_{dd}(T_{P}+E_{V})$&$10.0\pm0.6$&$9.2\pm0.3$  &$0.9995\pm0.0005$
\\
 $K^{+}K^{\ast-}$&$\lambda_{ss}(T_{V}+E_{P})$&$1.62\pm0.15$&$1.8\pm0.1$ &$0.977\pm0.003$
\\
 $K^{-}K^{\ast+}$&$\lambda_{ss}(T_{P}+E_{V})$&$4.50\pm0.30$&$4.3\pm0.2$  &$0.977\pm0.003$
\\
 $K^{0}\overline{K}^{\ast0}$&$\lambda_{ss}E_{P}+\lambda_{dd}E_{V}$&$0.18\pm0.04$&$0.19\pm0.03$  &$1.0\pm0.0$
\\
 $\overline{K}^{0}K^{\ast0}$&$\lambda_{ss}E_{V}+\lambda_{dd}E_{P}$&$0.21\pm0.04$&$0.19\pm0.03$  &$1.0\pm0.0$
\\
 $\eta\rho^{0}$&$\frac{1}{2}\lambda_{dd}(C_{V}-C_{P}-E_{P}-E_{V})\cos\phi-\frac{1}{\sqrt{2}}\lambda_{ss}C_{V}\sin\phi$&&$1.4\pm0.2$ &1
\\
 $\eta^{\prime}\rho^{0}$&$\frac{1}{2}\lambda_{dd}(C_{V}-C_{P}-E_{P}-E_{V})\sin\phi+\frac{1}{\sqrt{2}}\lambda_{ss}C_{V}\cos\phi$&&$0.25\pm0.01$ &1
\\
 $\pi^{0}\rho^{0}$&$-\frac{1}{2}\lambda_{dd}(C_{P}+C_{V}-E_{P}-E_{V})$&$3.82\pm0.29$&$4.1\pm0.2$ &1
\\
 $\pi^{0}\omega$&$-\frac{1}{2}\lambda_{dd}(C_{V}-C_{P}+E_{P}+E_{V})$&$0.117\pm0.035$&$0.10\pm0.03$ &1
\\
 $\pi^{0}\phi$&$\frac{1}{\sqrt{2}}\lambda_{ss}C_{P}$&$1.35\pm0.10$&$1.4\pm0.1$ &1
\\
 $\eta\omega$&$\frac{1}{2}\lambda_{dd}(C_{V}+C_{P}+E_{P}+E_{V})\cos\phi-\frac{1}{\sqrt{2}}\lambda_{ss}C_{V}\sin\phi$&$2.21\pm0.23$&$2.0\pm0.1$ &1
\\
 $\eta^{\prime}\omega$&$\frac{1}{2}\lambda_{dd}(C_{V}+C_{P}+E_{P}+E_{V})\sin\phi+\frac{1}{\sqrt{2}}\lambda_{ss}C_{V}\cos\phi$&&$0.044\pm0.004$ &1
\\
 $\eta\phi$&$\lambda_{ss}(\frac{1}{\sqrt{2}}C_{P}\cos\phi-(E_{P}+E_{V})\sin\phi)$&$0.14\pm0.05$&$0.18\pm0.04$ &1
\\
\hline
 $\pi^{0}K^{\ast0}$&$\frac{1}{\sqrt{2}}\lambda_{ds}(C_{P}-E_{V})$&&$0.103\pm0.006$ &$1.0\pm0.0$
\\
 $K^{0}\rho^{0}$&$\frac{1}{\sqrt{2}}\lambda_{ds}(C_{V}-E_{P})$&&$0.039\pm0.004$ &$1.0\pm0.0$
\\
 $\pi^{-}K^{\ast+}$&$\lambda_{ds}(T_{P}+E_{V})$&$0.345_{-0.102}^{+0.180}$&$0.40\pm0.02$ &$0.9994\pm0.0006$
\\
 $K^{+}\rho^{-}$&$\lambda_{ds}(T_{V}+E_{P})$&&$0.144\pm0.009$ &$0.983\pm0.002$
\\
 $\eta K^{\ast0}$&$\lambda_{ds}(\frac{1}{\sqrt{2}}(C_{P}+E_{V})\cos\phi-E_{P}\sin\phi)$&&$0.017\pm0.003$ &$1.0\pm0.0$
\\
 $\eta^{\prime}K^{\ast0}$&$\lambda_{ds}(\frac{1}{\sqrt{2}}(C_{P}+E_{V})\sin\phi+E_{P}\cos\phi)$&&$0.00055\pm0.00004$ &$1.0\pm0.0$
\\
 $K^{0}\omega$&$\frac{1}{\sqrt{2}}\lambda_{ds}(C_{V}+E_{P})$&&$0.064\pm0.003$ &$1.0\pm0.0$
\\
 $K^{0}\phi$&$\frac{1}{\sqrt{2}}\lambda_{ds}E_{V}$&&$0.024\pm0.002$ &$1.0\pm0.0$
\\
\hline \hline
\end{tabular}
\end{table}

Based on Eqs.~(\ref{eq:paraPP})
and (\ref{eq:paraPV}), we calculate the $D\to PP$ and $D\to PV$
contributions to $y$  by means of Eq.~\eqref{eq:yamp}, deriving
\begin{align}
y_{PP}=&~(1.00\pm0.19)\times10^{-3},\label{eq:yDPP}\\
y_{PV}=&~(1.12\pm0.72)\times10^{-3},\label{eq:yDPV}
\end{align}
respectively.
Our results are consistent with those in Ref.~\cite{Cheng:2010rv}:
$y_{PP}=(0.86 \pm 0.41)\times10^{-3}$, $y_{PV}=(2.69\pm2.53)\times10^{-3} ~(A,A1)$
and $y_{PV}=(1.52 \pm2.20)\times10^{-3} ~(S,S1)$ from two different solutions,
but with much smaller uncertainties. We stress that the predictions for $y_{PP}$
and $y_{PV}$ presented in this work are the most precise to date.
The uncertainties of the parameters in Eqs.~\eqref{eq:paraPP} and
\eqref{eq:paraPV} are basically controlled by those most precisely measured
channels, explaining why $y_{PP}$, with the
more precise $PP$ data, is more certain than $y_{PV}$. It is also the reason why the fit results for the most precisely 
measured branching ratios like $\mathcal{B}(\pi^0\overline{K}^0)$ and $\mathcal{B}(\overline{K}^0\omega)$ 
have uncertainties similar to those of the data, while the fit results 
for the less precisely measured ones like $\mathcal{B}(\eta\eta)$ and $\mathcal{B}(\pi^-K^{*+})$ 
have considerably smaller uncertainties. Besides, the branching
ratios are correlated to each other by the strong parameters
in the FAT approach, so the uncertainties are greatly reduced. 
Since the $SU(3)$ symmetry is assumed in the topological diagrammatic approach
\cite{Cheng:2010rv}, the charm mixing parameter $y$ cannot be extracted in
principle. Instead, the data of the branching ratios were directly input into
Eq.~\eqref{eq:ybr} by taking $\cos\delta_{n}=1$ \cite{Cheng:2010rv} as mentioned
before, such that the uncertainties of the data are summed up in the evaluation of $y$.
Some other efforts have been devoted to global fits of the $PP$ or $PV$ data
recently \cite{Muller:2015lua,Biswas:2015aaa,Cheng:2016ejf}. However, it
is unlikely that a precise prediction for $y$ can be made without thorough exploration
of the $SU(3)$ breaking effects in the relevant $D$ meson decays.

\section{$y_{VV}$}

There exist three different polarizations in the final state
of a $D\to VV$ channel, whose corresponding amplitudes can
be expressed in the transversity basis ($A_0$, $A_{||}$, $A_\perp$), or
equivalently in the partial-wave basis ($S$, $P$, $D$).
The decay amplitudes for different polarizations are independent, and
should be described by different sets of strong parameters in the FAT approach.
At least six strong parameters are required for the longitudinal amplitude
$A_0$ alone, but only one channel has been observed, with the longitudinal branching
ratio $\mathcal{B}_0(D^0\to\rho^0\rho^0)$ =(1.25$\pm$0.10)$\times10^{-3}$ \cite{PDG}.
The situation for the transverse amplitudes is even worse. It is impossible to extract all the $D\to VV$ amplitudes in the FAT approach
due to the lack of experimental data at present.

As a bold attempt, we estimate the $D\to VV$ longitudinal
amplitudes by means of the strong parameters in Eq.~(\ref{eq:paraPV})
extracted from the $PV$ data. In detail, the factorizable part in an emission-type
amplitude is treated in the naive factorization hypothesis,
and the associated nonfactorizable amplitude $\chi^C_{V}e^{i\phi^C_{V}}$
is assumed to be identical to that of the corresponding $PV$ amplitude.
We adopt the definition of the vector meson decay constant $f_V$ via
\begin{equation}
\langle V(q)|\bar{q}\gamma_\mu(1-\gamma_5)q'|0\rangle = f_{V}m_{V}\varepsilon^*_\mu(q),
\end{equation}
and the definition of the $D\to V$ transition form factors $V^{DV}$, $A_1^{DV}$,
$A_2^{DV}$, and $A_0^{DV}$ via
\begin{equation}\begin{split}
\langle V(k)|\bar{q}\gamma_{\mu}(1-\gamma_5)c|D(p)\rangle=&
\frac{2}{m_D+m_{V}}\epsilon_{\mu\nu\rho\sigma}\varepsilon^{*\nu}p^{\rho}k^{\sigma}V^{DV}(q^2)\\
&-i\bigg(\varepsilon^*_{\mu}-\frac{\varepsilon^*\cdot
q}{q^2}q_{\mu}\bigg)(m_D+m_{V})A_1^{DV}(q^2)\\
&+i\bigg((p+k)_{\mu}-\frac{m_D^2-m_{V}^2}{q^2}q_{\mu}\bigg)\frac{\varepsilon^*\cdot
q}{m_D+m_V}A_2^{DV}(q^2)\\
&-i\frac{2m_V(\varepsilon^*\cdot q)}{q^2}q_{\mu}A_0^{DV}(q^2),
\end{split}\end{equation}
where $\varepsilon$ is the polarization vector, the $m$'s are the meson masses,
and the momentum $q=p-k$. The emission-type amplitudes are then expressed as
\begin{eqnarray}
T(C)&=&{G_F\over\sqrt{2}}V_\text{CKM} a_1(\mu)\left( a_2^C(\mu)\right) f_{V_1}m_1\nonumber\\
& &\times\bigg[-ix(m_D+m_2)A_1^{DV_2}(m_1^2)
+i\frac{2m_D^2p_c^2}{(m_D+m_2)m_1m_2}A_2^{DV_2}(m_1^2)\bigg],
\end{eqnarray}
in which the Wilson coefficients and the kinetic quantities are given by
\begin{eqnarray}
& &a_1(\mu) = \frac{C_1(\mu)}{N_c} + C_2(\mu), \qquad
a_2^C(\mu) = C_1(\mu) + C_2(\mu)\left(\frac{1}{N_c}+\chi^C_{V} e^{i\phi^C_{V}}\right),\\
& &x=\frac{m_D^2-m_1^2-m_2^2}{2m_1m_2}, \qquad
p_c^2=\frac{m_1^2m_2^2(x^2-1)}{m_D^2},
\end{eqnarray}
respectively. The values of the form factors $A_{1,2}^{DV}$ are input from Ref.~\cite{Verma:2011yw}.
The annihilation-type amplitudes are taken directly from the $PV$
modes with the replacement of the meson masses and decay constants,
explicitly written as
\begin{equation}\begin{split}
E=&-i{G_F\over\sqrt{2}}V_\text{CKM}C_2(\mu)\chi^{E}_{q(s)}e^{i\phi^{E}_{q(s)}}
f_D{f_{V_1}f_{V_2}\over f_\rho^{2}}m_D|p_c|.\\
\end{split}\end{equation}

After estimating the $D\to VV$ longitudinal amplitudes, we can derive the
corresponding branching ratios straightforwardly. The comparison of our
predictions with the data will tell whether the $PV$-inspired amplitudes
are reasonable. The $D^{0}\to VV$ longitudinal branching ratios
in the FAT approach are listed in Table \ref{tableVV}, and compared with the
data of the total and longitudinal branching fractions. A general
consistency with the data is seen, especially for the single observed
longitudinal branching ratio $\mathcal{B}_{\rm long}(D^0\to\rho^0\rho^0)$.
For those channels with only measured total branching
ratios, most of our predictions for the longitudinal branching ratios do
not exceed the data, after considering the uncertainties.
Our result for the $D^{0}\to \overline K^{*0}\omega$ mode is larger than
the data, but the measurement of this mode was performed in 1992
\cite{Albrecht:1992tc}, and should be updated.
It is thus a fair claim that our simple estimates for the $D^{0}\to VV$
longitudinal amplitudes are satisfactory. Certainly, more experimental effort
toward improved understanding of the $D\to VV$ decays into final states
with different polarizations is encouraged.

A longitudinal amplitude $A_0$ is a linear combination of the partial waves
$S$ and $D$, namely, of the $L = 0$ and 2 final states, leading to $\eta_{\rm CP}(n) = +1$
in Eq.~\eqref{eq:yamp}. Inserting the amplitudes
estimated above into Eq.~\eqref{eq:yamp}, we obtain the longitudinal
$VV$ contribution
\begin{equation}
y_{VV} =(-0.42\pm0.34)\times10^{-3}.\label{yvv}
\end{equation}
The central value of $y_{VV}$ is lower than those of $y_{PP}$ and $y_{PV}$
in Eqs.~\eqref{eq:yDPP} and \eqref{eq:yDPV}, because the $SU(3)$
breaking effects are much smaller in the $VV$ modes. Even though Eq.~(\ref{yvv})
contains a relatively large uncertainty in our approach, and the
contributions from the transverse polarizations have not yet been included, it is
reasonable to postulate that $y_{VV}$ represents
a minor contribution to $y$.

In summary, our predictions for the mixing parameter $y$ agree well with
the postulation in Ref.~\cite{Gronau:2012kq}: $y$ is generated only at the second order
in $U$-spin symmetry breaking effects, so the contribution to $y$ from two-body
modes, for which the $U$-spin symmetry works
better, might be small. Multi-particle decays, for which the $U$-spin
breaking effects are expected to be more significant, should be
the major source of $y$.
We stress that we do not attempt a full understanding of $y$ here, and our
results for $y_{PP}$, $y_{PV}$ and $y_{VV}$ are consistent with the
fact that $y$ is generated at second order in the $SU(3)$
symmetry breaking.

\begin{table}[!tb]
\caption{Branching ratios for the $D^{0}\rightarrow VV$ decays in units of $10^{-3}$.
Estimations of the longitudinal branching ratios in the FAT approach are
compared with the data for the total and
longitudinal branching ratios \cite{PDG}.}\label{tableVV}
\begin{tabular}{cccccc}
\hline \hline
    ~~~~Modes~~~~  &~~~$\mathcal{B}_{\rm tot}$(exp)~~~&~~~$\mathcal{B}_{\rm long}$(exp)~~~&~~~$\mathcal{B}_{\rm long}$(FAT)~~~\\
\hline
$\rho^{0}\overline{K}^{\ast0}$&$15.9\pm3.5$&&13.2$\pm$1.3\\
$\rho^{+}K^{\ast-}$&$65.0\pm25.0$&&34.7$\pm$1.4\\
$\overline{K}^{\ast0}\omega$&$11.0\pm5.0$&&34.9$\pm$2.7\\
$\rho^{+}\rho^{-}$&&&3.2$\pm$0.1\\
$ K^{\ast+}K^{\ast-}$&&&1.1$\pm$0.05\\
$ K^{\ast0}\overline{K}^{\ast0}$&&&0.010$\pm$0.002\\
$\rho^{0}\rho^{0}$&$1.83\pm0.13$&$1.25\pm0.13$&1.1$\pm$0.1\\
$\rho^{0}\omega$&&&0.95$\pm$0.07\\
$\rho^{0}\phi$&&&0.65$\pm$0.04\\
$\omega\omega$&&&0.47$\pm$0.07\\
$\omega\phi$&&&1.41$\pm$0.09\\
$\rho^{0}K^{\ast0}$&&&0.038$\pm$0.004\\
$\rho^{-}K^{\ast+}$&&&0.123$\pm$0.005\\
$ K^{\ast0}\omega$&&&0.100$\pm$0.008\\
\hline \hline
\end{tabular}
\end{table}

\section{Summary}

In this paper we have calculated the \dd mixing parameter $y$ in the FAT approach,
considering the $D^{0}\to PP$, $PV$, and $VV$ channels. The $D^{0}\to PP$ and
$PV$ decay amplitudes were extracted from the latest data using the FAT approach, and the
$D^{0}\to VV$ decay amplitudes for the longitudinal polarization were estimated via the
parameter set for the $PV$ modes. It has been confirmed that the $PV$-inspired
amplitudes work well for explaining the observed $D^{0}\to VV$ branching ratios.
We then derived the contribution from the $PP$ and
$PV$ modes as
\begin{align}
y_{PP+PV}=(0.21\pm0.07)\%,
\end{align}
which is much more precise than previous predictions in the literature, and
far below the data $y_{\rm exp}=(0.61\pm0.08)\%$. It has been also found
that the contribution from the longitudinal $VV$ modes, being of order  $10^{-4}$,
is negligible. This observation is consistent with the
fact that $y$ is generated at second order in  $SU(3)$
symmetry breaking. We conjecture that considering the above two-body $D$ meson decays
alone in an exclusive approach cannot account for the charm mixing, and that
hadronic channels to other two-body and multi-particle final states are
relevant to the evaluation of $y$. However, it is currently very difficult, if not
impossible, to gain full control of the $SU(3)$ symmetry breaking
effects in all these modes in an exclusive approach.
As stated in the Introduction, the inclusive approach
leads to values of $x$ and $y$ two or three orders of magnitude lower than the
data. Therefore, a new strategy has to be proposed for
complete understanding of the charm mixing dynamics in the Standard Model.
We will leave this subject to a future project.

\section*{Acknowledgements}
The authors are grateful to Hai-Yang Cheng, Xiao-Rui Lyu, and Yang-Heng Zheng
for enlightening discussions. This work was supported in part by National
Natural Science Foundation of China under Grant No. 11347027, 11505083, 11375208, 11521505, 1162113100, 11235005 and U1732101,
by the MOST, Taiwan under
Grant No. MOST-104-2112-M-001-037-MY3, and by DFG Forschergruppe FOR 1873 ``Quark Flavour Physics and Effective Field Theories".


\clearpage

\end{document}